\documentstyle[12pt,epsf]{article}
\renewcommand{\baselinestretch}{1.5}

\begin{document}

\title{ASCA Measurements of Silicon and Iron Abundances 
in the Intracluster Medium}
\renewcommand{\baselinestretch}{1.0}

\author{
Yasushi {\sc Fukazawa},$^1$ Kazuo {\sc Makishima},$^1$ Takayuki {\sc Tamura},$^1$\\
Hajime {\sc Ezawa},$^1$ Haiguang {\sc Xu}, $^{1,2}$ Yasushi {\sc Ikebe},$^3$\\
Ken'ichi {\sc Kikuchi},$^4$ and Takaya {\sc Ohashi}$^4$ \\
{\small \it $^1$ Department of Physics, Graduate School of Science, 
              University of Tokyo,}\\
{\small \it   7-3-1 Hongo, Bunkyo-ku, Tokyo 113-0033}\\
{\small \it   E-mail(YF) fukazawa@miranda.phys.s.u-tokyo.ac.jp}\\
{\small\it $^2$ Institute for Space Astrophysics, Department of Physics,} \\
{\small\it    Shanghai Jiao Tong University,
              Huashan Road 1954, Shanghai 200030, PRC} \\
{\small \it $^3$ Max-Planck-Institut f\"ur Extraterrestrische Physik,}\\
{\small \it   Giessenbachstra\ss e, D-85748 Garching, Germany}\\
{\small \it $^4$ Department of Physics, Faculty of Science, 
              Tokyo Metroplitan University, }\\
{\small \it   Minami-Ohsawa, Hachioji, Tokyo, 192-0397}
}
\date{}
\maketitle

\renewcommand{\baselinestretch}{1.5}

\begin{abstract}

We analyzed the ASCA X-ray data of 40 nearby clusters of galaxies,
whose intracluster-medium temperature distributes in the range of 
0.9--10 keV.
We measured the Si and Fe abundances of the intracluster medium,
spatially averaging over each cluster,
but excluding the central $\sim 0.15 h_{50}^{-1}$ Mpc region in order
to avoid any possible abundance gradients and complex temperature structures.
The Fe abundances of these clusters are 0.2--0.3 solar, 
with only weak dependence on the temperature of the intracluster medium,
hence on the cluster richness.
In contrast, the Si abundance is observed to increase
from 0.3 to 0.6--0.7 solar from the poorer to richer clusters.
These results suggest that the supernovae of both type-Ia and type-II 
significantly contribute to the metal enrichment of the intracluster medium,
with the relative contribution of type-II supernovae 
increasing towards richer clusters.
We suggest a possibility that a considerable fraction of 
type-II supernova products escaped from poorer systems.
\end{abstract}


\section{Introduction}

The X-ray emitting hot intracluster medium (ICM) of clusters 
of galaxies is known to contain a large amount of heavy elements,
presumably processed in the stellar interior and ejected into 
the intracluster space (e.g. Hatsukade 1989; Arnaud et al. 1992; Tsuru 1992).
However, the mechanisms which supplied such a large amount of metals 
to intracluster space has not yet been understood well,
due to insufficient knowledge of the abundance ratios 
and spatial metallicity distributions in the ICM. 
In particular, relative importance of type-Ia supernovae 
(SNe Ia; e.g. Mihara, Takahara 1994) and type-II supernovae 
(SNe II; e.g. Arnaud et al. 1992) has remained controversial.
Since the SNe Ia products are iron-enriched, 
while the SNe II products are rich in $\alpha$-elements, 
such as O, Ne, Mg, and Si, 
we vitally need to make separate measurements of the 
Fe abundance and those of $\alpha$-elements.

Using the X-ray data acquired with ASCA (Tanaka et al. 1994), 
Mushotzky et al. (1996) studied abundance ratios of four clusters 
with the ICM temperature of $kT$=3--4 keV, and found 
that $\alpha$-elements in their ICM are roughly twice 
more abundant than Fe in solar abundance units.
This was also confirmed by Tamura et al. (1996) and Xu et al. (1997).
Mushotzky et al. (1996) hence suggest 
that SNe II are the dominant source of the metals in the ICM.
However, a study of hotter (richer) and cooler (poorer) clusters 
remains to be performed.

We present here a summary of ASCA measurements of the Fe and Si 
abundances in the ICM of 40 nearby clusters with various degrees of richness.
The Hubble constant is expressed as $H_0$=50 $h_{50}$  km s$^{-1}$ Mpc$^{-1}$.
The solar abundances refer to the solar photospheric values 
by Anders and Grevesse (1989), 
with (Fe/H)$_{\odot}=4.68\times10^{-5}$ 
and (Si/H)$_{\odot}=3.55\times10^{-5}$.

\section{Observations and Data Reduction}

We selected our sample clusters from the ASCA archival data 
through the following rather loose criteria:
the object must be extended sufficiently to resolve spatially; 
it must be brighter than $\sim 1\times10^{-12}$erg s$^{-1}$ cm$^{-2}$ 
in 0.5--10 keV, so that the metal abundances can be well determined; 
and it must not exhibit any outstanding morphological peculiarity or asymmetry. 
In addition, we attempted to pick up clusters with various degrees of richness. 
Through these criteria we selected 40 clusters, as summarized in table 1.
These objects turned out to be nearby ones at redshifts of $z<$ 0.062,
with the ICM temperature spanning a range of 0.9--10 keV.

These objects were observed with the GIS (Gas Imaging Spectrometer; 
Ohashi et al. 1996; Makishima et al. 1996) operated in the normal PH mode,  
and the SIS (Solid-state Imaging Spectrometer) 
operated mostly in the 4CCD Faint/Bright mode. 
We screened data requiring the cut-off rigidity to be $>8$ GeV c$^{-1}$ 
and the target elevation angle above the earth rim to be $>5^{\circ}$. 
In the SIS data selection we used event grades of 0, 2, 3, and 4,
and further required the elevation angle 
above the day earth rim to be $>25^{\circ}$.  

Clusters of galaxies often exhibit cool emission components 
and abundance increases near to the center (Fukazawa et al. 1994; 
Fabian et al. 1994; Matsumoto et al. 1996; Xu et al. 1996; Ikebe et al. 1997). 
Such a region should be excluded from the present study,
because we wish to focus on the average ICM properties. 
We therefore accumulated the GIS and SIS spectra 
for each cluster over a ring-shape region,
which typically has an inner radius of 0.1$h_{50}^{-1}$Mpc 
and an outer radius of 0.4$h_{50}^{-1}$Mpc from the cluster center.
For each cluster, we then added all of the available SIS data 
from  different sensors (SIS 0 and SIS 1), chips, and modes,
into a single SIS spectrum after an appropriate gain correction.
Similarly, data from GIS 2 and GIS 3 were summed 
into a single GIS spectrum for each object.

Utilizing $\sim 100$ ks data from several blank-sky observations,
we accumulated background events for each cluster.
We used the same data-integration region 
and the same data-selection criteria
as were used for the on-source data integration.
The derived background spectra were then subtracted
from the corresponding on-source spectra.
Figure 1 gives blow-ups of the Si-K line regions of the SIS spectra,
where individual spectra from clusters with similar temperatures, 
redshifts, and photon statistics have been co-added for presentation.

\section{Data Analysis and Results}

\subsection{Spectral Fits}
For each individual cluster, we fitted the background-subtracted GIS and SIS 
spectra jointly, with the variable-abundance single temperature (1T) 
Raymond-Smith model (R-S model: Raymond, Smith 1977).
To avoid any residual uncertainties in the instrumental response 
we limited the SIS and GIS fittings to energy ranges of
0.65--9.0 keV and 1.1--10.0 keV, respectively. 
The abundances of He, C, and N were fixed at the solar values,
while the heavier elements were divided into five groups: 
O and Ne; Mg; Si; S, Ar, and Ca; and Fe and Ni.
We constrained the elements in the same group 
to have a common abundance in solar units.
The free parameters of the fit were the Galactic interstellar absorption 
($N_{\rm H}$),
temperature ($kT$), normalization, and abundances of the five element groups. 

The fit has mostly been acceptable with a reduced chi-square 
of $<1.3$ for a typical degree of freedom of 200.
No significant discrepancies were observed between the two instruments.
Thus, the spectra of most clusters can be represented 
fairly well by the 1T R-S model.
A few very bright objects gave a somewhat larger reduced chi-square as 1.5--1.8,
possibly due to the residual uncertainties in the response matrices and the 
Fe-L line modeling (Fabian et al. 1994).
However, the best-fit values are thought to be usable.

Table 1 summarizes the results of these spectral fits.
Thus, the spatially averaged Fe and Si abundances were 
determined with a typical accuracy of 10\% and 40\%, respectively.
We plot the derived Fe and Si abundances in figures 2a and 2b, 
respectively, as a function of the ICM temperature.
Other elemental abundances remain poorly constrained.

\subsection{Iron Abundances}

In figure 2a, the obtained Fe abundance distributes over 0.2--0.4 solar, 
without any strong temperature dependence.
This abundance range roughly reconfirms the previous results 
using non-imaging instruments (e.g. Hatsukade 1989; Tsuru 1992).
However, the negative dependence of the Fe abundance on the temperature
(hence on the richness),
observed with Ginga (Hatsukade 1989; Tsuru 1992), is absent in figure 2a.

This discrepancy might arise because the ASCA results 
for $kT<3$ keV rely strongly upon the Fe-L lines,
which are known to be problematic (Fabian et al. 1994).
(The Ginga measurements rely solely upon the Fe-K line,
because of the harder bandpass of the Ginga LAC.)
To examine this issue, we analyzed the spectra of clusters with $kT>1.7$ keV,
ignoring either the Fe-K line region or the Fe-L line region of the spectrum.
However, we did not find any systematic discrepancy beyond 10\%
between the abundances determined with these two sets of Fe lines.
Similar confirmation has been reported by Mushotzky et al. (1996) 
and Hwang et al. (1997).

For a further study, in figure 2c we compare the R-S fitting results 
with those employing plasma emission codes by Masai (1984) 
and by Mewe-Kaastra-Liedahl (Mekal: Mewe et al. 1985; Liedahl et al. 1995): 
R-S, Masai, and Mekal codes are known to differ in the 
Fe-L line treatment (Fabian et al. 1994; Fukazawa et al. 1996).
In figure 2c, we have also averaged the individual abundance measurements 
over five broad temperature intervals.
The ensemble-averaged Fe abundance is thus amazingly constant
over a temperature range $kT=$ 1.5--7 keV.
Furthermore, we can see a very good agreement 
among the three codes above $kT \sim 1.5$ keV,
although in cooler clusters they become discrepant due to the Fe-L problem.
We conclude that the Fe-L determination is reliable, 
except in the poorest objects.

Given these arguments, the richness-dependent Fe abundances derived with Ginga
are likely to be affected by the metal concentration near to the cD galaxies, 
because they refer to the emission-weighted values.
Actually, when the central cluster regions are retained in the ASCA analysis,
the Fe abundances become negatively dependent on $kT$ (Fukazawa 1997).
We hence infer that the Fe abundances of poor clusters
increase towards the central region.
Apart from the central region, the ring-shaped regions used in the present 
study were further subdivided, 
for each object, into finer spatial regions by Fukazawa (1997)
in search for metallicity variations on finer spatial scales.
However, none was found beyond typical upper limits of 30--50\%.

Another interesting feature seen in figure 2c is the apparent 
abundance drop in the hottest clusters.
However, these results are subject to systematic uncertainties 
in the temperature measurements, 
because $kT \sim 10$ keV is somewhat too high for the ASCA bandpass.
Furthermore, these rich clusters tend to exhibit 
irregular variations in temperature and abundance. 
Therefore we regard this result as tentative.

\subsection{Silicon Abundances}

Figure 2b provides the first comprehensive report about the Si abundances of the intracluster medium (ICM).
The measured Si abundance is 0.6--0.7 solar in rich clusters ($kT>4$ keV),  
in agreement with the previous ASCA measurements (e.g. Mushotzky et al. 1996).
Moreover, in contrast to Fe, the Si abundance appears to correlate 
positively with the ICM temperature.
This tendency can be more clearly seen in figure 2d
in the form of ensemble-averaged values,
where the three plasma codes are confirmed to agree with one another.
However, the Si-K lines make much weaker spectral features
than the well studied Fe-K lines.
In addition, the Si-K lines fall close to the instrumental
Si-K edge of the SIS and the Au-M edge of the XRT.
These urge us to critically examine the reliability 
of the Si abundance measurements.

We accordingly closely inspect the ensemble-averaged spectra of figure 1, 
where the best-fit 1T R-S models with the Si abundance reset to 0 are superposed.
The H-like Si-K lines (at 2.01 keV in the rest frame) are clearly seen 
in panels (a) through (d) of figure 1, at equivalent widths (EWs) of 
$\sim$ 60--70 eV, $\sim$ 40--50 eV, $\sim 20$ eV, and $\sim 10$ eV, respectively.
When combined with the continuum temperature, these EWs respectively
yield Si abundances of $\sim 0.3$, 0.4--0.5, $\sim 0.6$ and $\sim 0.7$ solar,
in agreement with those derived with the 1T R-S fit (figure 2d).
In comparison, an analysis of the SIS/GIS spectra of bright AGNs indicates
that any false line in the Si-K line energy regions,
arising as an instrumental artifact, is $< 4$ eV in EW.
Therefore, the measured Si abundances are subject to systematic errors
by no larger than 20\% at $kT<4$ keV, and 30\% at $kT=6$ keV.
Even considering these systematic errors, 
the temperature dependence of the Si abundance remains real, 
at least below $kT \sim 6$ keV.

A more serious concern is 
that the Si lines in hot clusters might be contaminated 
by emission from any co-existing cool component, 
which would emit the Si-K lines at much larger EWs.
To examine this possibility,
we jointly fitted the ensemble-averaged SIS (figure 1c)
and GIS (not shown) spectra of clusters with $kT \sim 4$ keV,
with a model involving two R-S components (2T R-S model).
The two components are allowed to have free normalizations and temperatures,
but are constrained to have common absorption and common abundances.
As shown in table 2, the fit was somewhat improved over the 1T R-S model,
and the Si abundance decreased by $\sim 0.1$ solar.
Nevertheless, the Si abundance is still significantly higher
than those in cooler clusters ($\sim 0.3$; figure 2b, figure 2d).
The allowed contribution from the cool component (with $kT \sim 1.5$ keV; table 2)
is at most 10\% in the energy range of the Si-K lines;
a larger amount of the cool component would emit too strong Fe-L lines
to be allowed by the observed spectra.
If we tie the Si and Fe abundances together in the 2T R-S fit, 
the fit chi-square increases by 17.8.

As a cross confirmation, we utilize the fact 
that the cooler two spectra in figure 1 
exhibit He-like  Si-K lines (at a rest-frame energy of 1.86 keV)
in addition to the H-like ones.
Then, we should also observe He-like Si-K lines in the spectra of 
hotter clusters, if they were contaminated by cool emission.
To quantitatively examine this possibility,
we again fitted jointly the ensemble-averaged SIS/GIS spectra
of clusters with $kT \sim 4$ keV, 
using the 1T R-S model over a narrow energy band of 1.6--3.0 keV.
Then, the temperature is expected to become
sensitive to the H-like/He-like line ratios.
As shown in table 2, this analysis yielded a temperature range 
which is somewhat lower than that obtained in the wide-band 1T fit.
Although the Si abundance decreased to $\sim 0.54$, 
it is still significantly higher than those found in clusters with $kT \sim 1$ keV.
Similarly, we fit only the hard energy band (3.0--10.0 keV), 
the obtained temperature increases
by only 0.3 keV than that obtained by the wide band fitting (see table 2).
This also confirms that the emission is approximately isothermal.

From these examinations, we concluded 
that the temperature dependence of the Si abundance is real, 
at least over the ICM temperature range of $kT<6$ keV,
although the true dependence may be somewhat weaker 
than is indicated by figure 2d.
Finally, in figure 3, we present the ensemble-averaged Si to Fe 
abundance ratios in solar units as a function of the ICM temperature.

\section{Discussions}

We have measured spatially averaged Si and Fe abundances 
of the ICM of 40 nearby clusters, excluding the central regions.
Our results concerning the Si abundance confirm the previous report by 
Mushotzky et al. (1996), and extend it to a much larger sample,
in that the Si/Fe ratio is high at 1.5--2 (in solar unit)
in clusters with considerable richness.
This indicates that SNe II play a significant role in hotter clusters,
because such abundance ratios differ from those found in the SNe Ia products
(however see also Ishimaru, Arimoto 1997).

We have also discovered that the Si/Fe abundance ratio 
decreases towards poorer clusters.
If the metals are produced by the SNe II alone,
this demands the chemical composition of the SNe II products 
to depend on the cluster richness, due, e.g., 
to the difference in the initial mass function of stars.
This is, however, unlikely,
since no difference has been seen between poorer and richer clusters
in terms of the color-magnitude relation of member ellipticals
(Visvanathan, Sandage 1977).
Dust confinement of silicon in poor clusters is not likely either, 
because the dust evaporation time scale is at most $10^8$ yr 
in the ICM environment (Itoh 1989).

Thus, the varying Si/Fe ratio may not be explained without invoking 
increasing contributions from SNe Ia in poorer clusters.
This inference is reinforced by comparing clusters with supernova remnants (SNRs):
the X-ray spectra of very poor ($kT\sim1$ keV) clusters (e.g. Fukazawa et al. 1996) 
resemble very much those of type-Ia SNRs showing strong Fe-L lines,
while distinct from those of type-II SNRs exhibiting prominent O-K, Ne-K, 
and Mg-K lines (Hayashi et al. 1994; Hughes et al. 1995).
Furthermore, Ishimaru and Arimoto (1997) discuss 
that more than half the Fe in the ICM can be produced by SNe Ia, 
even in rich clusters.
We therefore conclude that both types of supernovae contribute 
significantly to the metal enrichment of the ICM,
with the relative contribution of SNe Ia increasing towards poorer clusters.
This provides the first observational evidence that SNe Ia contribute 
significantly to the metal enrichment of the ICM.

The richness-dependent relative contributions from the 
two types of SNe, in turn,  may allow two alternative explanations.
One is to invoke a richness dependence in their relative frequencies.
This may be possible, because richer clusters clearly exhibit higher 
fractions of elliptical galaxies than do poorer clusters (Dressler 1980),
and ellipticals are thought to have had higher frequencies of SN II
per unit mass than spirals in the past (e.g. Arimoto, Yoshii 1987).
However, the recent HST discovery of rapid morphological evolutions of member 
galaxies in rich clusters is against this possibility.
(Dressler et al. 1994).

Alternatively, the observed variation in the Si/Fe ratio may be explained
if the SNe II products have higher specific energies than the SNe Ia products.
Then, a larger fraction of SNe II products will escape from
a shallower gravitational potential of poorer systems,
while the SN Ia products may be confined even by the poorest systems,
thus producing the observed trend in the Si/Fe ratio.
This is quite plausible, 
because the SNe II products must have been supplied mostly in the 
form of galactic wind during early phase of cluster formation, 
whereas SNe Ia occurred on much prolonged time scales as isolated events.
We expect this effect to have been even stronger during the galactic wind phase,
when the gravitational potential may well have been shallower than today.
Actually, there is good evidence that a large quantity of supernova products
escaped from individual elliptical galaxies (Matsushita 1997), 
and to a less extent, from groups of galaxies (Fukazawa et al. 1996).
Further investigation of the metal confinement efficiency 
will be presented in a forthcoming paper.

\vspace*{0.8cm}
The authors are grateful to Prof. K. Yamashita and Prof. Y. Tanaka
for valuable comments, and to the ASCA team for their help 
in the spacecraft operation and calibration.

\clearpage
\section*{References}

\begin{list}{}{\setlength{\leftmargin}{3em}\setlength{\rightmargin}{
0cm}
	\setlength{\itemsep}{0.6ex}\setlength{\baselineskip}{-0.1ex}
	\setlength{\itemindent}{-3em}}

\item Anders E., Grevesse N. 1989, Geochim. Cosmochim. Acta. 53, 197
\item Arimoto N., Yoshii Y. 1987, A\&A 173, 23
\item Arnaud M., Rothenflug R., Boulade O., Vigroux L., Vangioni-Flam E. 1992, A\&A 254, 49
\item Dressler A. 1980, ApJ 236, 351
\item Dressler A., Oemler A., Sparks W.B., Lucus R.A. 1994, ApJ 435, L23
\item Fabian A.C., Arnaud K.A., Bautz M.W., Tawara Y. 1994, ApJ 436, L63
\item Fukazawa Y., Ohashi T., Fabian A.C., Canizares C.R., Ikebe Y., Makishima K., Mushotzky R.F., Yamashita K. 1994, PASJ 46, L55
\item Fukazawa Y., Makishima K., Matsushita K., Yamasaki N., Ohashi T., Mushotzky R.F., Sakima Y., Tsusaka Y. et al. 1996, PASJ 48, 395
\item Fukazawa Y. 1997, Ph.D. Thesis, The University of Tokyo
\item Hatsukade I. 1989, Ph.D. Thesis, Osaka University
\item Hayashi I., Koyama K., Ozaki M., Miyata E., Tsunemi H., Hughes J.P., Petre R. 1994, PASJ 46, L121
\item Hughes J.P., Hayashi I., Helfand D., Hwang U., Itoh M., Kirshner R., Koyama K., Markert T.  et al. 1995, ApJ 444, L81
\item Hwang U., Mushotzky R.F., Loewenstein M., Markert T.H., Fukazawa Y., Matsumoto H. 1997, ApJ 476, 560
\item Ikebe Y., Makishima K., Ezawa H., Fukazawa Y., Hirayama M., Honda H., Ishisaki Y., Kikuchi K. et al. 1997, ApJ 481, 660
\item Ishimaru Y., Arimoto N. 1997, PASJ 49, 1
\item Itoh H. 1989, PASJ 41, 853
\item Liedahl D., Osterheld A., Goldstein W. 1995, ApJ 438, L115
\item Makishima K., Tashiro M., Ebisawa K., Ezawa H., Fukazawa Y., Gunji S., Hirayama M., Idesawa E. et al. 1996, PASJ 48, 171
\item Masai K. 1984, Ap\&SS 98, 367
\item Matsumoto H., Koyama K., Awaki H., Tomita H., Tsuru T., Mushotzky R,F., Hatsukade I. 1996, PASJ 48, 201
\item Matsushita K. 1997, Ph.D. Thesis, The University of Tokyo
\item Mewe R., Gronenschild E.H.B.M., van den Oord G.H.J. 1985, A\&AS 62, 197
\item Mihara K., Takahara F. 1994, PASJ 46, 447
\item Mushotzky R.F., Loewenstein M., Arnaud K.A., Tamura T., Fukazawa Y., Matsushita K., Kikuchi K., Hatsukade I. 1996, ApJ 466, 686
\item Ohashi T., Ebisawa K., Fukazawa Y., Hiyoshi K., Horii M., Ikebe Y., Ikeda H., Inoue H. et al. 1996, PASJ 48, 157
\item Raymond J.C., Smith B.W. 1977, ApJS 35, 419
\item Tamura T., Day C.S., Fukazawa Y., Hatsukade I., Ikebe Y., Makishima K., Mushotzky R.F., Ohashi T. et al. 1996, PASJ 48, 671
\item Tanaka Y., Inoue H., Holt S.S. 1994, PASJ 46, L37
\item Tsuru T. 1992, Ph.D. Thesis, The University of Tokyo
\item Visvanathan N., Sandage A. 1977, ApJ 216, 214
\item Xu H., Ezawa H., Fukazawa Y., Kikuchi K., Makishima K., Ohashi T., Tamura T. 1997, PASJ 49, 9

\end{list}

\small
\begin{table}[htbp]
\caption[Cluster]{Sample clusters of galaxies.}
\begin{center}
\begin{tabular}{cccc|cccc}
\hline
\hline
Target & $kT$ & Si & Fe & Target & $kT$ & Si & Fe \\
       & (keV) & (solar) & (solar) &       & (keV) & (solar) & (solar) \\ 
\hline
Ophiuchus \dotfill & 10.26$\pm$0.32 & 1.06$\pm$0.41 & 0.21$\pm$0.02 & 	Centaurus \dotfill & 3.68$\pm$0.06 & 0.91$\pm$0.21 & 0.44$\pm$0.03 \\ 
Tri.Aust.$^{\ast}$ \dotfill & 10.05$\pm$0.69 & 2.26$\pm$1.22 & 0.19$\pm$0.06 & 	Hydra-A  \dotfill & 3.57$\pm$0.10 & 0.56$\pm$0.24 & 0.23$\pm$0.04 \\ 
A2319    \dotfill & 8.90$\pm$0.34 & 0.46$\pm$0.54 & 0.17$\pm$0.03 & 	A1367    \dotfill & 3.55$\pm$0.08 & 0.43$\pm$0.19 & 0.19$\pm$0.03 \\ 
Coma     \dotfill & 8.38$\pm$0.34 & 0.78$\pm$0.43 & 0.24$\pm$0.03 & 	A539     \dotfill & 3.24$\pm$0.09 & 0.52$\pm$0.23 & 0.18$\pm$0.04 \\ 
A2256    \dotfill & 7.08$\pm$0.23 & 0.97$\pm$0.39 & 0.24$\pm$0.03 & 	A1060    \dotfill & 3.24$\pm$0.06 & 0.36$\pm$0.12 & 0.30$\pm$0.02 \\ 
A478     \dotfill & 6.90$\pm$0.35 & 0.92$\pm$0.45 & 0.21$\pm$0.03 & 	2A 0335+096 \dotfill & 3.01$\pm$0.07 & 0.51$\pm$0.13 & 0.30$\pm$0.03 \\ 
Perseus  \dotfill & 6.79$\pm$0.12 & 0.67$\pm$0.18 & 0.38$\pm$0.02 & 	A194     \dotfill & 2.63$\pm$0.15 & 0.37$\pm$0.31 & 0.25$\pm$0.09 \\ 
A3571    \dotfill & 6.73$\pm$0.17 & 0.94$\pm$0.34 & 0.25$\pm$0.02 & 	Virgo    \dotfill & 2.58$\pm$0.03 & 0.49$\pm$0.06 & 0.34$\pm$0.02 \\ 
A85      \dotfill & 6.31$\pm$0.25 & 0.73$\pm$0.42 & 0.29$\pm$0.03 & 	AWM 4     \dotfill & 2.38$\pm$0.17 & 0.24$\pm$0.25 & 0.33$\pm$0.10 \\ 
A1795    \dotfill & 5.88$\pm$0.14 & 0.62$\pm$0.28 & 0.26$\pm$0.03 & 	A400     \dotfill & 2.31$\pm$0.14 & 0.44$\pm$0.22 & 0.31$\pm$0.13 \\ 
A119     \dotfill & 5.59$\pm$0.27 & 1.04$\pm$0.83 & 0.24$\pm$0.06 & 	MKW 9     \dotfill & 2.23$\pm$0.13 & 0.64$\pm$0.54 & 0.40$\pm$0.11 \\ 
A3558    \dotfill & 5.12$\pm$0.20 & 0.58$\pm$0.38 & 0.21$\pm$0.04 & 	A262     \dotfill & 2.15$\pm$0.06 & 0.37$\pm$0.15 & 0.27$\pm$0.05 \\ 
A2147    \dotfill & 4.91$\pm$0.28 & 1.05$\pm$0.38 & 0.31$\pm$0.06 & 	MKW 4s    \dotfill & 1.95$\pm$0.17 & 0.45$\pm$0.27 & 0.25$\pm$0.08 \\ 
A496     \dotfill & 4.13$\pm$0.08 & 0.60$\pm$0.18 & 0.31$\pm$0.03 & 	MKW 4     \dotfill & 1.71$\pm$0.09 & 0.44$\pm$0.18 & 0.33$\pm$0.07 \\ 
A2199    \dotfill & 4.10$\pm$0.08 & 0.73$\pm$0.19 & 0.30$\pm$0.03 & 	NGC 507   \dotfill & 1.26$\pm$0.07 & 0.28$\pm$0.13 & 0.33$\pm$0.07 \\ 
A4059    \dotfill & 3.97$\pm$0.12 & 0.61$\pm$0.25 & 0.39$\pm$0.04 & 	HCG 51    \dotfill & 1.23$\pm$0.06 & 0.23$\pm$0.11 & 0.31$\pm$0.06 \\ 
AWM 7     \dotfill & 3.75$\pm$0.09 & 0.49$\pm$0.19 & 0.33$\pm$0.03 & 	Fornax   \dotfill & 1.20$\pm$0.04 & 0.32$\pm$0.13 & 0.23$\pm$0.03 \\ 
A2634    \dotfill & 3.70$\pm$0.28 & 0.65$\pm$0.32 & 0.24$\pm$0.10 & 	NGC 5044  \dotfill & 1.07$\pm$0.01 & 0.27$\pm$0.07 & 0.27$\pm$0.02 \\ 
MKW 3s    \dotfill & 3.68$\pm$0.09 & 0.62$\pm$0.23 & 0.27$\pm$0.03 & 	HCG 62    \dotfill & 1.05$\pm$0.02 & 0.09$\pm$0.08 & 0.15$\pm$0.03 \\ 
A2063    \dotfill & 3.68$\pm$0.11 & 0.79$\pm$0.28 & 0.23$\pm$0.04 & 	NGC 2300  \dotfill & 0.88$\pm$0.03 & 0.53$\pm$0.48 & 0.15$\pm$0.07 \\ 
\hline
\multicolumn{8}{l}{$\ast$: Triangulum Australis cluster}
\end{tabular}
\end{center}
\end{table}

\end{document}